\documentstyle[12pt,cites,epsf,rotate,amssymb,amsfonts]{article}
\begin{document}

\title{\bf Evangelos Anastassakis:\\Scientist, Colleague and Friend}
\author{Manuel Cardona\footnote{m.cardona@fkf.mpg.de,~
 Fax: +49-711-689-1712}
\\[18pt]
Max-Planck-Institut f\"ur Festk\"orperforschung, \\
Heisenbergstr.\ 1, 70569 Stuttgart, Germany}
\date{}
\maketitle

\section{Evangelos, the Friend}
I first met Evangelos at the University of Pennsylvania in the
Fall of 1969. Fred Pollak and I had started setting up a Raman
laboratory at Brown University and, given Fred's interest and
experience in uniaxial stress techniques, we had decided to launch
a program for measuring the effect of such stress on Raman phonons
in solids, in particular semiconductors. We decided to begin with
silicon. The pressure and sample preparation equipment was readily
available at Brown but we had to wait for the Raman equipment and
the funds to pay for it (which we got, in due course, from the
University Administration!). At some conference we met Eli
Burstein, with whom I had a long friendship dating back to 1958 (I
was in 1958 a graduate student and Eli already a big man in Solid
State Physics. I was impressed ever since by his way of treating
me as an equal). Evangelos, then called by us "Van", may well have
been present at the meeting. Eli suggested that what we should do
is bring prepared samples and the pressure rig (we had to watch
out not to call it "bomb" so as to avoid a {\em faux pas} at the
airports) and perform the measurements in Philadelphia using his
hand-made Raman spectrometer (built by a highly skilled technician
named Aron Filler). We agreed and Fred and I went to Philadelphia
and worked with "Van" and Aron Pinczuk day and night, pressing Eli
into giving us as much time as he could as a busy professor.

I believed that one round of measurements of a few days would be
sufficient to collect the data needed to write the first
experimental paper on the matter. It took two rounds for Evangelos
to write the first draft. It was not the first paper Evangelos had
written. He had received his Ph.D. at Penn in 1968 and had already
published or sent for publication eight manuscripts. He knew very
well the ins and outs of Eli's spectrometer (very important when
dealing with homemade instruments!) and pitched in day and night
to try to finish the project as soon as possible.

The only anecdote I remember from those days is his leaving the
side of the instrument (those days such instruments were seldom
left to take data by themselves) to answer a phone call. He spoke
Greek and sounded very upset. All I was able to understand is his
repeated expletive: "katastrophia". The call was long and we were
quite concerned about his getting rather pale. When he hung up and
we asked for the reason, he said that a friend of his in Greece
had just been arrested (remember that the military had taken over
in Greece in 1967 and, I presume, in 1969 they were going through
a rather virulent phase). I tried later, whenever I met "Van", to
get details about the conversation but, as often in these cases
(remember the recently made public Heisenberg-Bohr conversation of
1941), there was a nebulous cloud about it. Not even Marilena has
been able to help. Be it as it may, Evangelos recovered from the
phone call and the five of us (Evangelos, Aron, Eli, Fred, and
myself, in that order) wrote the final version of a paper which
appeared, of course, in {\it Solid State Communications}
\cite{ana1}. The subject was considered, at that time, rather
esoteric and we did not expect that our paper would be seminal to
many other publications, both theoretical and experimental, to the
present date. With 236 citations over the years it became
Evangelos's most cited publication and it gave a permanent imprint
to his work. The paper has been cited this century already 20
times, having kept its impact 30 years after its having appeared.
This became possible, of course, by the application of the then
measured dependence of phonon frequencies on stress to the
characterization of semiconductor nanostructures.

In 1971 Evangelos became an Assistant Professor at Northeastern
University in Boston. In the face of the tough competition from
Harvard, MIT, Boston University and other institutions in the
area, his possibilities at Northeastern in terms of students and
equipment, were limited. Characteristic of his activities in the
years to follow, he looked for the equipment he needed elsewhere,
where his expertise was usually received with open arms
(outsourcing, as we call it today). So he spent the years 1973 and
1974 on and off as a visiting scientist at MIT, being promoted to
Associate Professor at Northeastern in 1973. In spite of the
limitations at that institution, I counted 30 publications which
can be attributed to the Northeastern period. They cover a wide
range of subjects, including infrared and Raman spectroscopy of
lunar materials obtained in the Apollo 11,12,14 and 15 missions
\cite{perry2}. This work has been cited 43 times.

In November of 1973 the Military Dictator was toppled (I avoid
mentioning his name so it does not appear as a citation in the
Citations Index \cite{cites}) after a series of bizarre events
which still today affect the Eastern Mediterranean. In July 1974
\underline{Karamanlis} (he deserves to be cited!) became President
and the long spook was over. Evangelos began to dream about a
return and chose our Institute in Stuttgart as a vantage point
from where to follow the events in Greece. I offered him a
position in my group in the hope that he may stay here for a long
haul, possibly permanently: I needed him rather badly. He arrived
in Stuttgart in September of 1974 and plunged into the laboratory
in spite of its being moved to a new site. But alas, his home
country needed to rebuild the badly battered universities. He was
appointed Associate Professor at the Technical University and
given travel orders to start teaching in Athens as of January 1,
1975. For me, it was a real "catastrophe", but he promised to come
regularly to spend the summers in Stuttgart, which he did. The
last collaborative effort before his untimely death was a 220
pages long article in ``Semiconductors and Semimetals''
\cite{ana4} on the effects of uniaxial stress on phonons, the
topic we had started in 1970.

One anecdote stuck in my mind from the short three-months period
of his stay at the end of 1974. He came in the morning and showed
me an infection, possibly staphylococci, in one of his fingers. It
was rather ugly, the kind we call in the Catalan language {\it un
voltadits}. We were just a five minute walk from the main
hospital. I suggested that he go there and have them look at it.
He did not return, so I called the hospital the next day and they
told me he had spent the night there as an in-patient, was
operated on, and was now waking up from ``total anesthesia''.  I
could not believe my ears. I went right away to the hospital and
they told me that they had strict orders to operate all Greeks
only under ``total anesthesia''. The reason, they claimed, is that
Greeks, like all Mediterraneans, have a tendency to faint at the
sight of blood and they did not want to take any chances. I have
seen a lot of blood in my life and have never fainted!

I would like here to quote from the first letter which I received
from Evangelos after his return to Greece, dated Jan. 30, 1975:

``Well, I made it all the way to Athens without any difficulty. My
infection is also gone. Classes have already started; I have two
2-hour floor-shows (I mean lectures) every week. My audience
consists of $\sim$ 1000 freshmen!! I feel like a political
candidate. They are basically well-behaving, very much aware of
what is going on, and extremely politicized. In addition to
teaching, there is a lot to be done in running their labs,
recitation meetings, and trying to reconcile the always
complaining personnel.''

My reply to him announced the positive solution of a serious
bureaucratic problem. When he came in September of 1974 from the
US, we had payed his transatlantic transportation. Before leaving
in December he had been told by someone in our administration that
he had to return the fare because the law required for the refund
a minimum stay with us of six months. This was a lot of money
those days and the Greek government had apparently no legal means
to pay the fare (unbelievable when you consider the cost of
appointing and supporting  a professor who is going to be active
for at least 25 years!). So I wrote a strong letter to our
administration in support of Evangelos, asking them to waive the
refund with all kinds of more or less specious arguments. In a
hand-written, informal, five lines note, the head of our
administration answered me that there was no problem whatsoever.
The corresponding paragraph of the law (a minimum stay of
six-months for the refund) applied only to relatives, not to the
visiting scientist. At least this time, German bureaucracy was
allowed to act in a rational way.

I could fill a few more pages with anecdotes, some sweet, some
sour. I remember we both having been invited to Wolfgang Richter's
apartment on the occasion of his birthday in June of 2000. We left
the party together, rather late, for our hotels and there were
some seedy looking characters in the area. Evangelos insisted in
delivering me to the door of my hotel. He was rather concerned
with my personal safety: a premonition of his untimely death only
a couple of months later? I remember very vividly what we talked
about: Emmanuel Royidis and his famous 19th century novel. Those
of you who know him may remember that Evangelos had an
encyclopedic culture. We had discussed often the dearth of modern
Greek literature till the end of the 19th century. He introduced
me to Royidis and his raucously irreverent novel. I tried long to
find a translation, to no avail. In January of 2000 I found one in
a used book store in California \cite{pope}. By the time we last
met in June, I had read it, so we had plenty of material to
discuss. We spent about an hour at the door of the Hotel
Heidelberg (the weather was mild) till he parted forever. He also
recommended me to read the ''Murderess'', by A. Papadiamantis.  I
just found a German translation \cite{murder} and am in the
process of reading it. Unfortunately, I will not be able to
discuss it with Evangelos this time ...

Evangelos untimely death left a big hole in the lives of many of
us. First of all, I must think about his wife Marilena, then about
the Technical University and even myself, who had known him since
1969, and collaborated with him regularly and steadily throughout
those long years. But I do not think that Evangelos would like us
to sit, wail and bemoan his death. He would implore that the
moaning be left, like in the Greek antiquity, to professional
wailers.

I would like here to celebrate Evangelos LIFE, to thank the powers
that each of us believes in for having made him possible, and to
celebrate and recall his important contributions to science.

\section{Evangelos, the Scientist}
As already indicated, Evangelos scientific life can be divided
into four, partly overlapping periods:
\begin{enumerate}
\item The University of Pennsylvania (1962-1968).
\item Northeastern University (1969-1974).
\item The Max-Planck-Institute for Solid State Research in
Stuttgart (1974 and a number of stays during the summers
thereafter).
\item The National Technical University of Greece (1975-2000).
\end{enumerate}

\noindent I shall present in what follows some highlights of his
scientific work during those periods, the choice being guided by
the criteria:
\begin{itemize}
\item What I consider to be the importance of the work.
\item The citations the publication has received, according to
a version of the ISI Web of Science (Institute of Scientific
Information, Philadelphia) which encompasses the beginning of
Evangelos career around 1965 (this version actually goes back to
1946!, so it even encompasses the beginning of Eli Burstein's
career, except for two papers published by Eli in 1940 \cite{bur7}
and 1941 \cite{bur8}).
\item The work in which I have collaborated most intensely.
\end{itemize}

\subsection{The University of Pennsylvania (Penn) (1962-1968)}
The first and one of the most important publications of Evangelos
\cite{ana9}, was based on work performed at Penn. It was
co-authored by Eli Burstein, his thesis advisor, and Sato Iwasa, a
bright Japanese graduate student who got his Ph.D. at Penn, also
under Eli, in 1966. Sato moved then to the corporate sector and
after a distinguished 25-years career with Lockheed and Honeywell,
working mainly on infrared detectors and optical systems, founded
his own company (IRtech Consultants, Harvard, Massachusetts) of
which today he is still president.

Reference \cite{ana9}, a short letter, has been cited 32 times.
However, to this number one must add 30 citations to a full-sized
paper on the same topic published in 1970 \cite{ana10}. I shall
discuss next the main ideas behind this important and seminal
work.

In 1965, laser Raman scattering was beginning to establish itself
as a powerful technique for the investigation of elementary
excitations in solids, in particular lattice vibrations (phonons).
The technique was, at the time, mainly used to measure the
frequency of such excitations (the so-called Raman shift
\cite{raman}). Little attention was being paid to the scattering
efficiency $S$, (defined as the ratio of the number of scattered
photons per unit path length and unit solid angle to the number of
incident photons). Only crude theoretical estimates already
existed \cite{lou}. Evangelos first paper \cite{ana9} led to the
first experimental determination of $S$ for a solid (diamond). It
was an indirect method, based on a thermodynamic type of argument,
and the numbers obtained have received ample confirmation in
later, more direct work [13-17]. When comparing the dates of Ref.
\cite{ana1}, (1966), with those of Refs. [13-17], (1970-1981,) we
realize how far ahead of his time Evangelos was in 1966. This
repeated itself a number of times later, especially in his work on
piezo-Raman spectroscopy \cite{ana1}.

The ``thermodynamic'' argument used by Evangelos in \cite{ana9}
had been suggested by Burstein and Ganesan in 1955 \cite{bur18},
but had to wait for two enterprising graduate students,
Anastassakis and Iwasa, to bring it to fruition  (Eli has also
often been ahead of his time!). Phonon Raman scattering takes
place through modulation of the electrical polarizability by the
time-dependent amplitude of the phonon, as represented by the
parameter {\it a} [with the dimensions of a length] which
describes the derivative of the polarizability with respect to the
phonon displacement. For the sake of simplicity, we shall here
neglect the tensor properties of that derivative and represent it
simply by the scalar {\it a} (given usually in $\rm\AA^2$). The
scattering efficiency $S$ for diamond structure materials is then
given by \cite{ana1}:

\begin{eqnarray} 
S~ =~r \frac{\hbar\omega_S^4N^2}{\varrho\omega_0 c^4} ~|a|^2 ~(n_o
+ 1) ,
\end{eqnarray}

\noindent where $r$ is a number of the order of one which
expresses the scattering selection rules, $\omega_S$ the scattered
frequency, $N$ the number of primitive cells per unit volume
$\varrho$ the crystal density, $\omega_0$ the phonon frequency and
$n_o$ the Bose-Einstein statistical factor ($n_o\simeq 0$ for
diamond at 300K. According to (1) a determination of $a$ leads to
the determination of $S$ and vice versa (except for the phase of
$a$ which in the region of transparency can only be zero or $\pi$,
i.e., $a$ can be either positive or negative). The so-called Raman
tensor $a$ can be written as (dropping tensor indices):
\begin{eqnarray} 
a~=~ \frac{\partial P}{\partial u}~ ,
\end{eqnarray}

\noindent where $P$ is the polarizability and $u$ the sublattice
displacement corresponding to the Raman phonon under
consideration. Diamond has two atoms per primitive cell (PC),
i.e., two sublattices. The phonon that corresponds to the
displacement of one sublattice with respect to the other is Raman
active, i.e., (2) does not vanish for it (there are actually three
such phonons, corresponding to displacements along the three
crystallographic directions). The middle point of the positions of
the two atoms in a PC is a center of inversion that leaves the
crystal invariant: it is said that the corresponding phonons are
{\it even} under inversion. Phonons can also be detected by
infrared absorption spectroscopy: The absorption coefficient
$\alpha$ can be written as:
\begin{equation} 
\alpha~ {\rm proportional~ to} ~\left|\frac{\partial M}{\partial
u}\right|^2
\end{equation}

\noindent where $M$ is the electric dipole moment of a finite
piece of the crystal. Since $\partial M/\partial u$ must be
invariant with respect to symmetry operations and $M$ is {\it odd}
with respect to the inversion while $u$ is even, $\alpha$ vanishes
for the Raman phonons of diamond (this is a rather general rule:
for any centrosymmetric crystal, allowed Raman phonons are
infrared forbidden, and vice versa). An electric field (odd
symmetry) applied to diamond breaks the inversion symmetry and can
make, in the absence of other symmetry restrictions, Raman phonons
ir-allowed. The corresponding contribution to (3) is:

\begin{eqnarray} 
\Delta\alpha \propto ~ \left|\left(\frac{\partial^2M}{\partial
u\partial E_0}\right)\right|^2 E_0^2 ~.
\end{eqnarray}

\noindent Considering that $(\partial M/\partial E_0) = P$, the
term in brackets in (4) turns out to be precisely the ``Raman
tensor'' $a$ of (2). Hence, a measurement of the ir absorption
induced in diamond at the Raman phonon frequency by an electric
field $E_0$ yields a value for $a$ and, using (1), for the
scattering efficiency $S$. To measure this effect, the directions
of $E_0$ and the ir field $E_{ir}$ must be judiciously chosen so
as to avoid a vanishing of Eq.(4) imposed by symmetry selection
rules. Evangelos chose $E_0$ along the [001] crystallographic
direction and $E_{ir}$ along [1$\bar{1}$0]: it is easy to figure
out that these directions yield the maximum $\Delta\alpha$ for a
given $E_0$.

Figure~1 shows typical experimental results for the effect of an
electric field on the ir transmission of diamond, as obtained by
Evangelos. The light path under the applied field, $E_0
=1.2\times10^5$ V/cm, was 2.4 mm. From the decrease of 1\% in
transmission seen in Fig.~2, a value of $|a|=3.8 \rm\AA^2$ was
determined in \cite{ana9}. Note that these measurements, as well
as those of \cite{ana10} and [13-17] do not yield information
about the sign of $a$ which is believed to be positive (the
longitudinal polarizability increases when the C-C bond is
stretched) on the basis of calculations \cite{grim17}.

\begin{figure}[hbt]
\begin{minipage}[t]{11.5cm}
 \centerline{\epsfxsize=4cm\epsffile{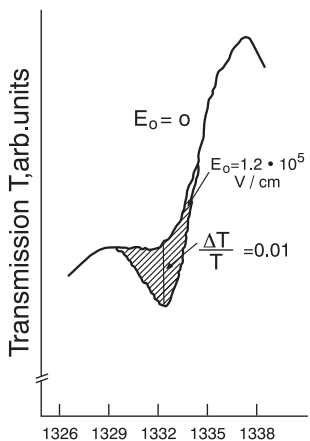}}
\caption{{\it Recorded transmission spectrum for a diamond with a
path length of 2.4 mm under zero and nonzero electric fields. The
relative change of the transmittance is about 1\% for $E_0 =
1.2\times 10^5$ V/cm [10]. See text.}} \label{cardfig1}
\end{minipage}
\end{figure}

\begin{figure}[hbt]
\begin{minipage}{11.5cm}
\centerline{\epsfxsize=9cm\epsffile{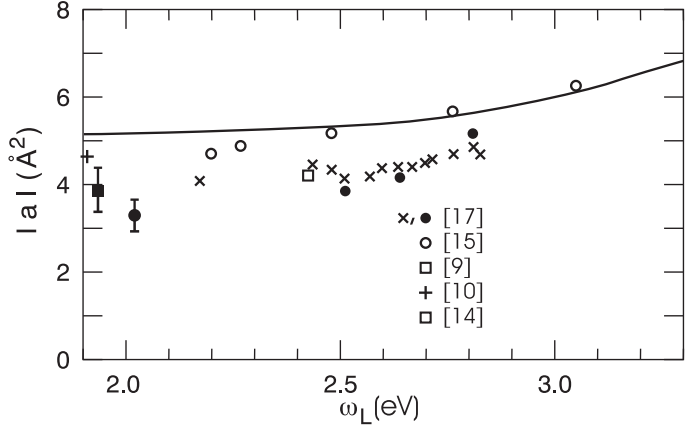}}
\caption[]{{\it Compilation of measurements of the Raman tensor
and its dispersion for diamond. The solid line represents the
equation $a (\rm\AA)^2 = -6.5$g $(\omega_L$/5.6 eV) discussed in
[17]. According to this equation, $a$ should be {\it positive}.}}
\label{cardfig2}
\end{minipage}
\end{figure}

The apex of experimental virtuosity was reached by Evangelos in
\cite{ana10} when he estimated, by using ac-techniques and lock-in
amplifiers, the effect of order $E_0^4$ on $\Delta\alpha$ and that
of order $E_0^2$ on the Raman tensor.

The second most important piece of work of the ``Penn Period'' is
probably contained in \cite{ana1}. We reproduce the unprocessed
experimental data in Fig.~3 so as to emphasize their quality,
especially considering the rather unsophisticated equipment used.
\begin{figure}[bth]
\begin{minipage}[b]{11.5cm}
\centerline{\epsfxsize=10cm\epsffile{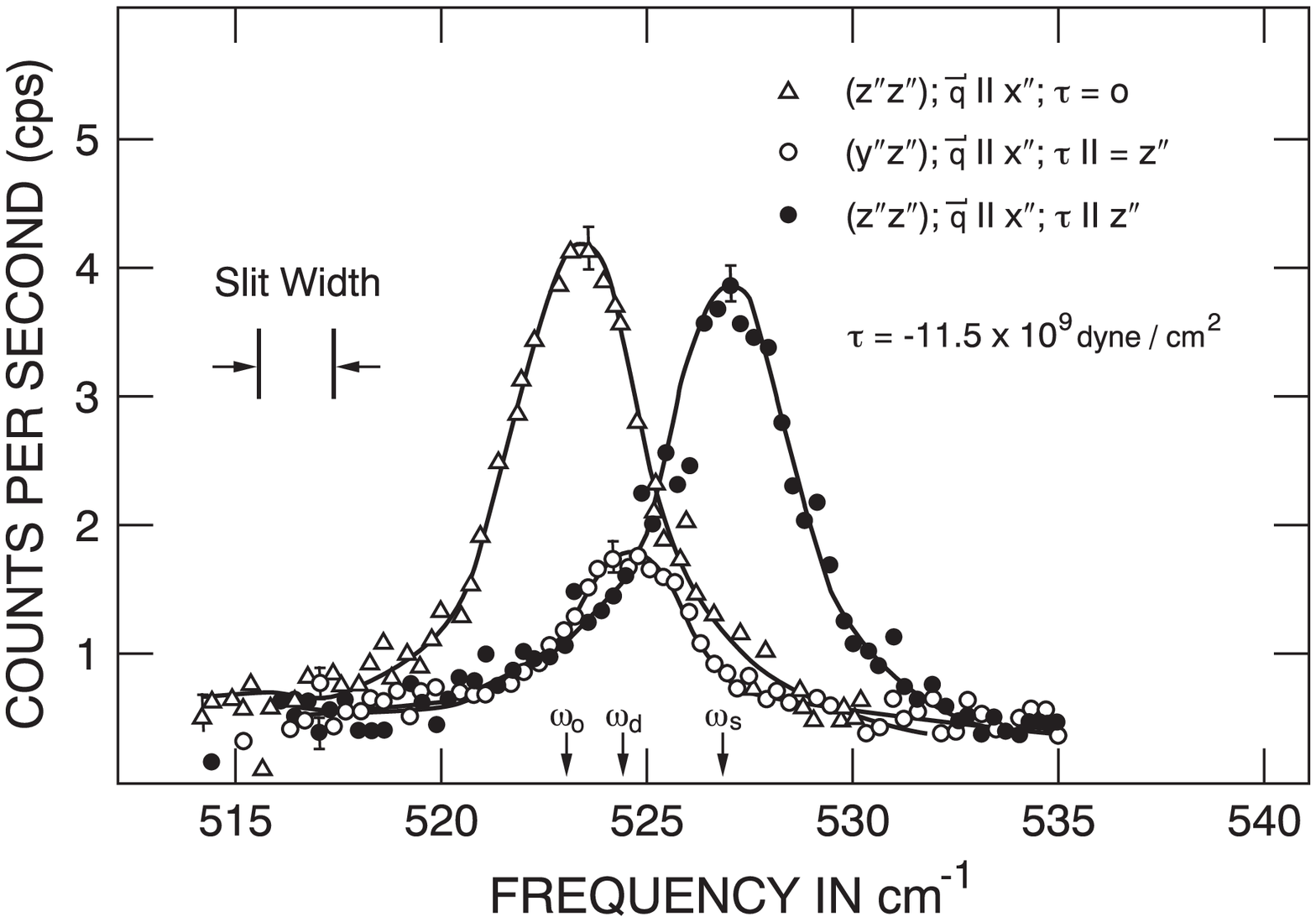}}
\end{minipage}\hfill
\begin{minipage}[b]{12cm}
\caption{{\it The first-order Raman spectrum of Si at
300$^{\circ}$K with a uniaxial stress $\tau = -11.5\times 10^9$
dynes/cm$^2$ applied along the [111] direction \cite{ana1}. The
entries in the parentheses, i.e. ($z''\| [111]$) and ($y''\|$
[$1\bar{1}0]$), designate the polarization direction of the
incident and scattered radiation. In each case, the direction of
the incident and scattered radiation is along $x''$ ($x''\|
[11\bar{2}$]) and $-x''$, respectively \cite{ana1}.}}
\label{cardfig3.}
\end{minipage}
\end{figure}

Figure~3 shows spectra of the Raman phonon of silicon both,
unstressed and under a {\it compressive} stress of $11.5\times
10^9$ dyne/cm$^2$ (1.15 GPa). Notice that under stress two peaks
are seen, depending on the directions of incident and scattered
light. Evangelos and I have given ever since to students the
problem of figuring out which is the direction of vibration of the
Raman phonons which correspond to either one of the two peaks. The
answer good students find is that the peak obtained for incident
and scattered polarizations parallel to the direction of the
stress ([111] direction) corresponds to phonons which vibrate
along that direction. For the other peak they vibrate
perpendicular to it. A good student will soon add that this
explains the direction in which the frequency shift takes place:
if you compress along the direction of vibration the ``springs''
holding the atoms are compressed. They therefore get stiffer and
the frequency should increase. This is correct for a compression
along [111] but, as found in \cite{ana1}, for a compression along
[100] the opposite takes place: the vibration along [100] {\it
softens} while the perpendicular ones harden. It takes a
\underline{very good} student, such as Evangelos was in those
days, to figure out the answer: A pure shear compression along
[100] does not change the length of the bonds since it
\underline{decreases} it by an amount $\Delta e$ along [100] and
\underline{increases} it by half as much along [010] and [001]. So
far so good, but then the good student goes and measures diamond,
as Evangelos did years later, when he was no longer a student
\cite{grim19}. He, working in Stuttgart with Marcos Grimsditch
\cite{argen} and myself, found that for diamond the longitudinal
vibrations (parallel to the stress) increase in frequency for a
compression either along [111] or [100]. Evangelos figured out a
simple way to explain that: \cite{grim19} diamond is much more
covalent than silicon and therefore the restoring forces causing
the lattice vibrations are not only strong for \underline{bond
stretching}, but also for \underline{bond bending}. It takes a
back-of-the-envelope calculation, of the type Evangelos excelled
in, to show that under these circumstances the longitudinal
vibrations also harden for a [100] compression \cite{card21}. He
also pointed out years later \cite{card21}, that the ``anomaly''
for a [100] compression also applied to boron nitride, a material
closely related to diamond.

Representative for the 20-odd papers related to the ''Penn
Period'', I would like to mention a series of five papers he
published under the general title of ``Morphic Effects, 1-5''
[22-26]. It was a nice way of summarizing his experimental work on
stress and field-effects on Raman phonons and presenting it under
a unifying group-theoretical formalism. The collection of 5 papers
has been cited a total of 120 times, a large number considering
the dryness of the rather formal subject.

I conclude the ``Penn Period'' by mentioning another seminal piece
of work having to do with electronic excitations in semiconductors
\cite{pinc}. Evangelos and co-workers observed electronic
excitations of an electron gas introduced by doping with donors
into bulk GaAs samples. Although this scattering is symmetry
forbidden, it becomes observable when the exciting laser frequency
is resonant with an electronic interband gap, in this case the
$E_0+\Delta_0$ gap at about 1.9 eV. The observed scattering
spectra depend rather strongly on whether the incident and
scattered polarizations are parallel or perpendicular. For
parallel polarizations, collective excitations, which reflect the
Coulomb interaction between the electrons, are observed. In the
perpendicular case, the scattering excitations involve a spin-flip
and thus are not affected by Coulomb interaction: one observes
single particle excitations.

\subsection{Northeastern University in Boston\\ 1970-1974}
The beginning of this period was spent in close collaboration with
his former mentor Eli and some of the corresponding work has
already been mentioned under 2.1. Evangelos work at Northeastern
was influenced by the collaboration with Clyve Perry, who, after a
varied and productive career in science and administration, is
still in the Faculty at Northeastern \cite{perry2,perry28}.

In \cite{perry28} work on ir-absorption and Raman scattering by
two magnons (plus a phonon for the ir-absorption) in NiO was
presented. In spite of the pioneer character of this work, it has
been cited only 12 times, a result of having been published in a
rather obscure journal (although it was part of a memorial issue
for C.V. Raman!).

To the Northeastern period belongs Evangelos brief excursion to
the moon. Reference \cite{perry2} has been cited about 50 times.
In it he collaborated with Perry and R.P. Lowndes, who around
1970, became Dean of the School of Arts and Sciences at
Northeastern,
 to remain in this position until 1997. The work describes a
detailed analysis of moon rocks by optical spectroscopies, through
comparison with similar rocks on earth. The authors concluded that
the lunar rocks are mainly composed of SiO$_2$, Al$_2$O$_3$, MgO,
and FeO, some of them containing significant quantities ($\sim
10\%$) of CaO and TiO$_2$.

Another paper belonging to this period was published in the
prestigious journal ``Science'' \cite{roth}. The work was a
collaboration among physicists and biochemists at Harvard, MIT and
Northeastern. Two of the coauthors, K.J. Rothschild  (a
biochemist) and Eugene Stanley (a statistical mechanics theorist)
are today highly reputable faculty members at Boston University.
I.M. Asher was connected with Harvard and with the Food and Drug
Administration Lab in Maryland. The paper describes the
application of Raman spectroscopy to the determination of the
structure of Valinomycin, an antibiotic. There was a follow-up to
this work: two longer papers in the Journal of the American
Chemical Society \cite{ash}. According to the byline, these papers
fall already well into the Athens Period. I have not been able to
figure out the logistics of such a collaboration.

As the last paper of this period, I mention his work with R.
Greenwald \cite{green}, an undergraduate student at Northeastern
who, following American custom (absent but much needed in
Europe!), moved on as a graduate student to the University of
Rochester. It contains beautiful work on optical properties of
crystals, in particular Bi$_{12}$GeO$_{20}$. This material, which
was rather topical at that time, has a cubic structure with the
point group T, the least symmetric of all five cubic
crystallographic point groups. As a consequence of the lack of
reflection planes, crystals belonging to this group are {\it
chiral} and exhibit optical rotatory power (i.e., they rotate the
plane of polarization of linearly polarized light). Greenwald and
Anastassakis measured the optical rotatory power of
Bi$_{12}$GeO$_{20}$ vs. wavelength and the effect of pressure and
temperature on this unusual but important property. The results
were interpreted on the basis of a single oscillator which was
allowed to depend on temperature and pressure. In spite of the
somewhat esoteric nature of Bi$_{12}$GeO$_{20}$, \cite{green} has
been cited 35 times.

\subsection{Athens}
As already mentioned, Evangelos moved in 1974 from Boston to the
Max-Planck-Institute in Stuttgart and, after three months,
continued to Athens. Until his death he kept also contact and
collaboration with my group in Stuttgart and also that of Wolfgang
Richter in Aachen and Berlin \cite{richt}. He had met Richter in
Philadelphia, where the latter was a post-doc with Eli, and met
him again in Stuttgart while Richter was working in my group: an
example of ``the web of science''. This section highlights the
work that Evangelos and his Greek associates performed in Athens.
Section 2.4 will discuss some of the work that resulted from his
collaboration with my group in Stuttgart.

When Evangelos arrived  to the National Technical University in
1975, he had no experimental facilities. Moreover, there were more
pressing duties than experimenting: he had to lecture to over 1000
students, as you read in Sec.~1, and he had to help repair the
ravages of seven years military dictatorship. Money for research
was nonexistent: he had to help develop in the political class the
awareness for the need of scientific, especially physics,
research: this may sound paradoxical for the country where physics
was invented. In order to stay fit, Evangelos began doing
paper-and-pencil work in Athens and traveling, during the
holidays, to some of the more fortunate friends who had
experimental facilities \cite{richt}. Slowly, but surely, he got
money to set up a modest lab in Athens where, by the early 1980's,
he had a Raman spectrometer and a diamond anvil cell to do optical
work under high hydrostatic pressure. By the mid 1980's
experimental work performed in his Athens lab began to appear in
the international literature. I would like to mention from this
period the work with Liarokapis \cite{lia33} in which they
measured the effect of intrinsic carriers, excited across the
small gap of InSb at high temperatures, on the phonon Raman
spectrum of InSb. Another publication, with Liarokapis and
Kourouklis \cite{lia34}, presents Raman data for LaF$_3$ vs.
temperature and pressure. The LaF$_3$ crystals, with rhombohedral
space group D$^4_{3d}$ under ambient conditions, display rather
unusual anharmonic behavior vs. pressure and temperature. Raman
measurements under pressures up to 10 GPa (100 kbar) were
performed in Athens at 300K and in Stuttgart, in a solid argon
environment, at 25K so as to check the consistency of the results.
I would have liked to share with Evangelos our recently gained
understanding of the temperature behavior of phonon self-energies
\cite{card35} and reinterpret with him the interesting data which
he published in \cite{lia34}.

More piezo-Raman work, in which Kourouklis and Evangelos
discovered a phase transition in SrF$_2$ at 300K, appeared in
\cite{kour}. In \cite{ana-rap} Anastassiadou, Raptis and Evangelos
discuss the effect of absorption and angular spread on Raman
scattering and compare the calculations with results they obtained
for GaAs, including the TO-LO splitting and the electrooptic
effect on the LO-scattering. Similar calculations had been
performed by the group of Miles Klein at the University of
Illinois, at Urbana \cite{klein} using a different approach. The
equivalence of the two methods is not obvious and should be
investigated.

Around 1986 Evangelos realized that Raman scattering was an
excellent technique for characterizing stresses and strains in
semiconductor nanostructures. He had actually laid the basis for
this work in \cite{ana1} and decided to apply his mastery of
crystal optics and tensor analysis to develop a general formalism
which could be used with as little computational effort as
possible. In \cite{ana39} Anastassakis and Liarokapis developed a
formalism for evaluating the splitting of phonons, equivalent to
that of Fig.~3, but for a polycrystalline sample. Considering the
strong anisotropy of such splitting, already mentioned in
connection with Fig.~3 and discussed in \cite{ana1}, complex
tensor averages must be performed. This work is a real ``tour de
force'' in tensor analysis of crystal properties. It has been
cited 40 times. Another highly cited publication \cite{ana40}
presents general expressions to calculate built-in strains at
heterojunctions of arbitrary origin due to lattice mismatch.
Evangelos and I were, at the time, unhappy about using the term
{\it biaxial} strains for isotropic strains in two dimensions. He
came up with the term {\it bisotropic} which he used consistently
ever since, as opposed to the commonly employed but misleadingly
incorrect designation of {\it biaxial}.

Going through the literature while writing this article, I
discovered a paper \cite{papa} in which a detailed but simple
modeling of the optical properties of diamond in the region of
electronic interband transitions (0-25 eV) was presented. I was
surprised to see that in it Evangelos has collaborated with {\it
Papadopoulos}, but soon I realized that it was A.D., not G.
Although this article has been cited 20 times, it was unknown to
me. Had I known it, I would have used it in some of the more
recent work I have done for diamond \cite{ruf}. My apologies,
Evangelos.

Last, but not least, I would like to mention the contribution of
Evangelos to the success of the memorable 20th International
Conference on the Physics of Semiconductors (20-ICPS), which was
held in Thessaloniki in August of 1990. The Proceedings of the
20-ICPS \cite{proc} were edited by Evangelos, together with J.D.
Joannopoulous, a well-known faculty member at MIT who has
distinguished himself recently by excellent work on photonic
crystals. I found in the citations index several papers from those
proceedings with a remarkable number of citations, contrary to
popular belief that conference proceedings are never cited. I
fear, however, that this ``popular belief'' may become a
self-fulfilling prophecy, if the present trend to electronic
publishing continues [see Proceed. of 24-ICPS (Jerusalem) and the
26-ICPS (Edinburgh)].

\subsection{Athens-Stuttgart Collaboration}
This fruitful collaboration lasted for a quarter of a century,
from 1974 until Evangelos death. The ``logistics'' was often that
he came to Stuttgart for a couple of months, collected data and
helped with their processing and the writing of a manuscript after
returning to Athens. The first such collaboration involved
measurements of the pressure dependence of the LO and TO
frequencies of GaAs with the diamond anvil cell \cite{trom}. It
was probably the first contact of Evangelos with this powerful
device. A mixture of methanol and ethanol was used in those days
as a pressure transmitting fluid, a pressure fluid which since has
been replaced by helium because of the problems involved in the
stiffening and freezing of alcohols under pressure. Nevertheless,
the results Evangelos obtained working with Rainer Trommer, a
German graduate student now at Siemens (see Fig.~4), stand today
as correct \cite{trom}. Once more against popular belief, the work
has been cited 50 times, in spite of having appeared in the
proceedings of a conference. The conference was held in Campinas,
Brazil, at the time of a rare economic boom in that country which
allowed the organizers to pay the air fare of basically all
participants (of course on VARIG, a Brazilian airline). I recall
that about 10 researchers from my group, including several
graduate students, belonged to this category, something that had
never happened before and probably will never happen again. I
believe that Evangelos was also there. This work confirmed our
conjecture that the effective charge $e^*$ decreases under
pressure in most semiconductors and led us to develop a
simple-minded explanation of this non-intuitive fact based on the
so-called {\it Laffer curve} (Fig.~5).

\begin{figure}[bth]
\begin{minipage}[b]{11.5cm}
\centerline{\epsfxsize=11.5cm\epsffile{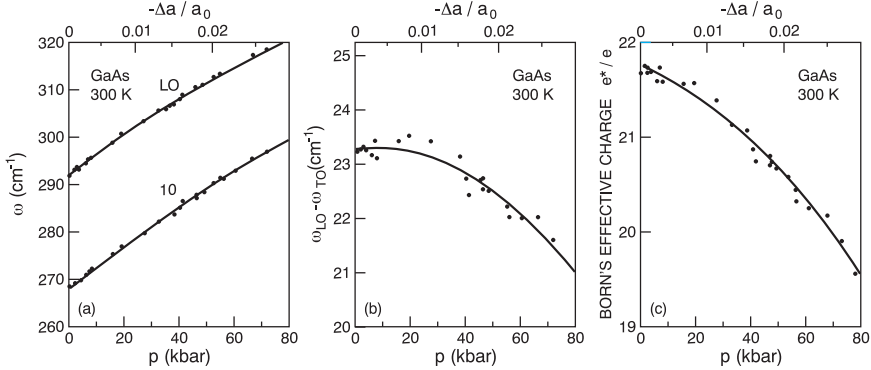}}
\end{minipage}\hfill
\begin{minipage}[b]{12cm}
\caption{{\it Effect of hydrostatic pressure on (a) the LO and TO
Raman frequencies of GaAs. (b) The difference between the
$\omega_{LO}$ and $\omega_{TO}$ frequencies. (c) Born's effective
charge vs. pressure as obtained from the data in (a) and (b)
\cite{trom}.}} \label{cardfig4.}
\end{minipage}
\end{figure}

This curve was allegedly suggested by Arthur B. Laffer at a dinner
party in 1974, drawing it on a napkin, in order to explain the
dependence of the state revenue on tax rate: if the tax rate is to
the right of the maximum, its increase will lead to a decrease in
revenue and vice versa (the people who believe this are called
``right wingers''). The opposite is true if the operating point is
to the left of the maximum. Laffer became economic advisor to
Ronald Reagan who, being a ``right winger'', decided on the basis
of Fig.~5 to lower the taxes so as to increase the internal
revenue of the US. The results of the experiment are well known.

\begin{figure}[bth]
\begin{minipage}[b]{11.5cm}
\centerline{\epsfxsize=6cm\epsffile{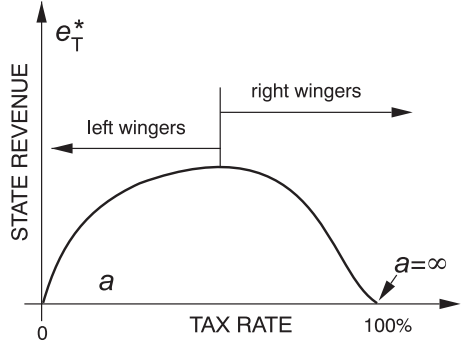}}
\end{minipage}\hfill
\begin{minipage}[b]{11.5cm}
\caption{{\it Curve proposed by Laffer ({\it Laffer's Curve}) to
represent the state revenue as a function of tax rate. It has been
used by Anastassakis and myself to represent the dependence of
Born's effective charge $e^*_T$ on lattice parameter $a$
(Fig.~4c). See text.}} \label{cardfig5.}
\end{minipage}
\end{figure}

In any case, the Laffer curve can also be taken to represent the
behavior of the ionicity of GaAs, i.e. $e^*_T$, vs. pressure. The
ionicity is zero for infinite lattice parameter since Ga as well
as As must then be neutral. It is also zero for a hypothetical
lattice parameter zero, since no charge transfer from Ga to As can
occur. Whether $e^*_T$ increases or decreases under pressure
(i.e., when the lattice parameter decreases) depends on the
position of the equilibrium lattice parameter with respect to the
maximum of the Laffer curve ...

Because of the already mentioned increasing interest in the
effects of uniaxial stress on phonons in semiconductors, pioneered
by Evangelos in \cite{ana1}, he took up in Stuttgart new aspects
of this problem, in particular those privy to ir-active phonons.
In \cite{wick} he collaborated with Paul Wickbold and Rolf Sauer
\cite{sauer}. Paul Wickbold had just finished his B.Sc. at Stony
Brook and was spending a few months in Germany, the land of his
ancestors, before going to work at Harvard with my former thesis
advisor, Bill Paul (the Web of Science again).

The splitting of TO and LO phonons of GaAs by a uniaxial stress
along [111] reported in \cite{wick} is shown in Fig.~6. An alert
student will understand why the TO phonons, a doublet, split. He
will, however, ask how come the LO phonon, a singlet, also splits.
A {\it very} alert student may find the answer: the LO and TO
phonons are not only characterized by the direction of {\it
vibration} but by that of {\it propagation}. The LO nomenclature
implies that the vibration is {\it parallel} to the propagation
direction, whereas it is perpendicular for the TO-phonons. Hence,
there are three degenerate LO phonons under zero stress: one which
vibrates and propagates along [111] and two which do both things
perpendicular to [111]. When a stress along [111] is applied, this
triplet splits into a singlet (propagating and vibrating along
[111] and a doublet (doing both things perpendicular to [111]).
This is expressed by the $s$ and $d$ subscripts in Fig.~6. The
splittings of both phonons in Fig.~6 correspond to the splittings
shown in Fig~1 for silicon but again, the alert student will note
that the $\omega_L$ and $\omega_T$ splittings are not equal in
Fig.~6. This was explained by Wickbold, {\it et al}. as reflecting
the anisotropic effect of the strain on $e^*_T$ for which a
phenomenological theory was developed in \cite{wick}. These
effects are nowadays calculated {\it ab initio} on the basis of
the electronic band structure \cite{card21}, nevertheless
phenomenological models can be very useful to trace errors such as
those pointed out in \cite{card21}. Reference \cite{wick} has
received 86 citations.

\begin{figure}[bth]
\begin{minipage}[b]{11.5cm}
\centerline{\epsfxsize=6cm\epsffile{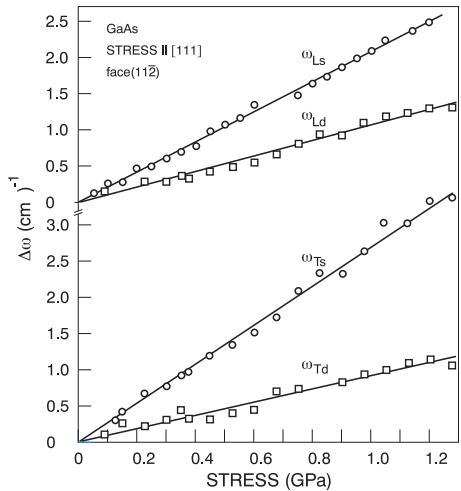}}
\end{minipage}\hfill
\begin{minipage}[b]{11.5cm}
\caption{{\it Splittings and shifts reported in \cite{wick} for
the TO and LO phonons of GaAs under a uniaxial stress along [111].
The subscripts $d$ and $s$ correspond to vibrations perpendicular
and parallel to the stress direction (see text).}}
\label{cardfig6.}
\end{minipage}
\end{figure}

Most of Evangelos publications on the subject of stress effects on
phonon frequencies have been highly cited. Belonging to the
Athens-Stuttgart period I would like to mention the work on
diamond \cite{grim19}, already discussed in Sect.~2.1, in
connection with \cite{ana1}. This work has been cited 110 times!

I would like to close the discussion of the Athens-Stuttgart
period by mentioning a paper \cite{ulr} which, in spite of being
rather recent (1997), has nevertheless been cited 21 times. This
paper belongs already to the era of {\it ab initio} calculations
but stands as a {\it caveat} to {\it ab initio} calculators that
not everything they calculate needs to be correct. Note the break
in the calculated linewidth vs. pressure at 7GPa, which is not
shown by the experimental data, a fact which was attributed in
\cite{ulr} to a slight error in the calculated frequencies of the
two-phonon decay channels which are responsible for the phonon
width. This error can be corrected by hand: the correction is
represented by the dashed-dotted line in Fig.~7. This discrepancy,
and the correction of the theoretical results so as to agree with
the experimental ones, is best expressed by a sentence which is
attributed to Max Planck, but could also have been Evangelos's:

``Experiments are the only means of knowledge to our disposal. The
rest is poetry, imagination.''

\begin{figure}[hbt]
\begin{minipage}[b]{12cm}
\centerline{\epsfxsize=5cm\epsffile{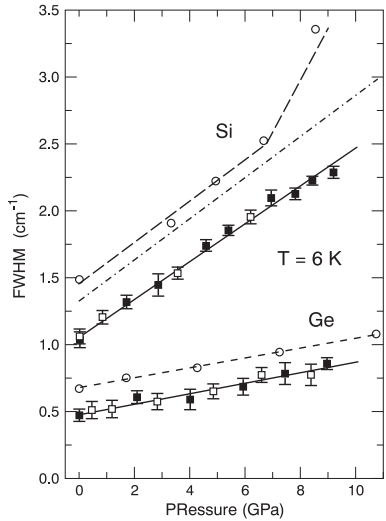}}
\end{minipage}\hfill
\begin{minipage}[b]{11.5cm}
\caption{{\it Measurements of the dependence of Raman phonon
linewidths of Si and Ge on pressure (symbols with error bars) as
reported in \cite{ulr}. The empty circles with dashed lines across
represent {\it ab initio} calculations: note the break for Si at
$\approx$ 7GPa which is an artifact of these calculations. The
dashed-dotted line represent the calculations after removal of the
artifact.}} \label{cardfig7.}
\end{minipage}
\end{figure}

\section{Conclusions}
In this article I have tried to portray Evangelos life as a human
being, a friend, and a scientist from the vantage point of our 32
years of friendship and collaboration. The Web of Science lists 26
papers jointly authored by Evangelos and myself. They are among
the most cited of his and my publications and, as such, many
appear in the enclosed list of references. I (and posthumously
Evangelos) would like to thank the many collaborators spread
around the world (our Web of Science) who made these publications
possible even if they have not been explicitly mentioned (many
have). His untimely death dealt a severe blow to Greek science. We
all miss him, but he will be mostly missed by his wife Marilena.
Let me thank her, in the name of us all, for the invaluable
support she gave him.


\begin{thebibliography}{47}
\bibitem{grim16} M.H. Grimsditch and A.K. Ramdas, Phys. Rev. B {\bf
                 11}, 3139 (1975). Here the efficiency for Raman scattering
                 was determined by comparison with that for Brillouin
                 scattering by acoustic phonons. The latter can be calculated
                 from the well-known elasto-optic constants.
\bibitem{sauer} R. Sauer is now a professor at the University of
Ulm which, Evangelos would agree, should better be named after
Albert Einstein, who was born in that town. After all, the
University of Thessaloniki is named after Aristotle.
\end{thebibliography}
\end{document}